# Onset of the Meissner effect at 65 K in FeSe thin film grown on Nb doped SrTiO$_3$ substrate

Zuocheng Zhang[1,4*], Yihua Wang[2,3*], Qi Song[2,5*], Chang Liu[1,4], Rui Peng[2,5], K.A. Moler[3], Donglai Feng[2,5 †] and Yayu Wang[1,4†]

[1]*State Key Laboratory of Low Dimensional Quantum Physics, Department of Physics, Tsinghua University, Beijing 100084, China*

[2] *State Key Laboratory of Surface Physics and Department of Physics, Fudan University, Shanghai 200433, China*

[3] *Stanford Institute for Materials and Energy Science, Stanford University, Stanford, CA 94305, USA*

[4]*Collaborative Innovation Center of Quantum Matter, Beijing 100084, China*

[5]*Collaborative Innovation Center of Advanced Microstructures, Fudan University, Shanghai 200433, China*

\* *These authors contributed equally to this work.*
† Email: dlfeng@fudan.edu.cn; yayuwang@tsinghua.edu.cn

**Abstract**: We report the Meissner effect studies on an FeSe thin film grown on Nb doped SrTiO$_3$ substrate by molecular beam epitaxy. Two-coil mutual inductance measurement clearly demonstrates the onset of diamagnetic screening at 65 K, which is consistent with the gap opening temperature determined by previous angle resolved photoemission spectroscopy results. The applied magnetic field causes a broadening of the superconducting transition near the onset temperature, which is the typical behavior for quasi-two-dimensional superconductors. Our results provide direct evidence that FeSe thin film grown on Nb doped SrTiO$_3$ substrate has an onset $T_C$ ~ 65 K, which is the highest among all iron based superconductors discovered so far.

The recent discovery of potential high temperature superconductivity in monolayer FeSe has attracted much attention in the high $T_C$ community [1]. The original report was a scanning tunneling microscopy (STM) study of monolayer FeSe film grown on SrTiO$_3$ (STO) substrate by molecular beam epitaxy (MBE) [2]. Tunneling spectroscopy revealed an energy gap as large as 20.1 meV [2], which is about an order of magnitude larger than the superconducting (SC) gap $\Delta = 2.2$ meV [3-5] for bulk FeSe with SC critical temperature $T_C \sim 9$ K [4, 6]. If we assume the same $2\Delta/k_B T_C$ ratio for the monolayer thin film and bulk samples, the estimated $T_C$ for the FeSe/STO system could be above 77 K [2]. It has been proposed that the unusually high $T_C$ is due to the strongly enhanced electron-phonon coupling strength at the FeSe/STO interface [2, 7-9], as well as the efficient transfer of electrons from the substrate to the FeSe film [10-12].

Soon after the original discovery, angle resolved photoemission spectroscopy (ARPES) measurements on similar FeSe/STO thin films confirmed the existence of an almost isotropic energy gap that persists to 65 K [13-15]. Both the momentum and temperature dependences of the energy gap are consistent with the behavior of SC gap in the K$_x$Fe$_2$Se$_2$ iron-based superconductors [13-15]. However, in order to confirm that the FeSe/STO system is indeed a high temperature superconductor with $T_C \sim 65$ K, one must demonstrate the existence of zero resistance and Meissner effect at similar temperature scales. Recently, *in situ* four probe transport measurements on FeSe/STO film showed the onset of zero resistance state near 100 K [9]. However, up to date the Meissner effect measurements on FeSe film have yet to show a convincing evidence for high temperature superconductivity. Previous Meissner effect measurement of monolayer FeSe grown on insulating STO substrate has shown a $T_C = 21$ K

[11].

In this letter, we report two-coil mutual inductance measurements on a FeSe thin film grown on Nb doped STO(001) substrate by MBE. The diamagnetic screening effect with an onset temperature of 65 K is clearly demonstrated, which is in agreement with the opening of a SC gap at the same temperature scale measured by ARPES [14, 15]. Magnetic field applied perpendicular to the film causes a broadening of the SC transition near the onset temperature rather than a parallel shift of $T_C$, which is consistent with the behavior of two-dimensional superconductors. Our results provide unambiguous evidence that FeSe thin film grown on Nb doped STO substrate has an onset $T_C$ ~ 65 K, which is the highest among all iron based superconductors discovered so far.

The FeSe film studied here is grown on a $TiO_2$ terminated and Nb doped (0.5 % wt) STO (001) substrate by using state-of-the-art MBE, and the treatment of substrate and deposition of FeSe film were carried out following the procedure described in Ref. [15]. For this sample we first grew a monolayer of FeSe on the substrate, and then annealed it at 600 °C for 3 hours in the ultra-high-vacuum MBE chamber. The post annealing process improves the sample quality and achieves the charge doping level that are crucial for the high temperature superconductivity. After that, we deposited two $(Fe_{0.96}Co_{0.04})Se$ layers, followed by two more FeSe layers. Finally an 18 nm thick amorphous Se layer was capped on top of the first FeSe layer to protect it from contamination or degradation when it is taken out of the MBE chamber. The schematic structure of the sample is shown in Fig. 1a.

Because the Nb doped STO substrate is highly conductive, it is difficult to directly measure the electrical transport properties of the FeSe thin film on the substrate. Therefore we focus on the Meissner effect measurement on this sample because the substrate makes negligible contribution to the diamagnetic signal. The two-coil mutual inductance measurement has been proved a highly sensitive method for detecting the existence of Meissner effect in SC thin films [11, 16, 17]. As illustrated in Fig. 1b, the sample is sandwiched between a pair of concentric coils. An ac current $I$ is applied through the drive coil to produce a changing magnetic flux $\Phi$, which will induce a voltage in the pickup coil. The pickup voltage $V_p$ is expressed as: $V_p = -N_p \frac{d\Phi}{dt} = -M \frac{dI}{dt}$. Here $N_p$ is the number of turns of the pickup coil, $M$ is the mutual inductance of the two coils, and $\frac{dI}{dt}$ is the changing rate of the current passing through the drive coil. In our experiments we use an ac drive current with amplitude $I_0 = 5$ μA and frequency $f = 10$ kHz. A standard ac lock-in method is used to measure the pickup voltage. When the sample becomes SC, the Meissner effect will screen the magnetic flux passing through the pickup coil, and thus the out-of-phase voltage $V_{out}$ decreases significantly [17].

The red curve in Fig. 2a displays the temperature dependence of $V_{out}$ across the pickup coil when the FeSe film on Nb doped STO substrate is inserted. At the high temperature regime (from 70 to 140 K), $V_{out}$ decreases very gradually with reducing temperature. This is mainly due to the Nb doped STO substrate, which becomes more conductive at low $T$ and provides slightly more efficient screening of the magnetic flux. In order to directly measure the background from the substrate, we perform the same mutual inductance measurement on a Nb doped STO substrate capped with 18 nm Se but without the FeSe film. The black curve in Fig. 2a shows such background signal, which is almost identical to the red curve at the high temperature regime. However, at around $T = 65$ K, the red curve starts to deviate from the black background curve, as indicated by the red arrow. This decrease of $V_{out}$ reveals the onset of diamagnetic screening, hence the Meissner effect, of the FeSe thin film. The $V_{out}$ signal drops rapidly below 40 K, and starts to saturate below 10 K. Figure 2b shows the derivative of the red curve shown in Fig. 2a, which clearly shows the onset of diamagnetic response from 65 K. This result is the first Meissner effect evidence showing that the FeSe film achieves macroscopic superconductivity below $T_c^{onset}$ = 65 K. This temperature scale is in excellent agreement with the ARPES measurements revealing the opening of SC gap at 65–70 K [14, 15].

We have also performed the mutual inductance measurements in varied magnetic field applied perpendicular to the film. Figure 3a are the temperature dependent of $V_{out}$ curves measured at $\mu_0 H$ = 0, 2, 5, 10, and 15 T respectively. The general trend is that with increasing magnetic field, the decrease amplitude of $V_{out}$ becomes smaller due to the suppression of superconductivity that leads to less efficient screening of the magnetic flux. However, the $V_{out}$ curves do not show a parallel shift towards lower temperature as expected for $T_C$ suppression by magnetic field in the mean field type behavior. Instead, they only spread out gradually starting from the same $T_c^{onset}$ = 65 K. In Fig. 3b we zoom into the onset regime of the same curves (marked by the dashed square in Fig. 3a), which clearly shows the broadening of the SC transition near 65 K. This behavior is typical for two-dimensional type II superconductors with strong fluctuations. Previous transport measurements on FeSe films grown on insulating STO substrate have also demonstrated similar two-dimensional SC properties [2].

It is worth noting that the $V_{out}$ drop shown in Fig. 2a is much smaller than that measured on previous FeSe film with the same sample size and even lower $T_C$ [11]. In the normal state at $T$ = 70 K, the $V_{out}$ values for both measurements are around 6.31 µV. At the base temperature $T$ = 1.5 K, the $V_{out}$ value for this sample is decreased to 5.56 µV. But for the previous FeSe film on insulating STO with $T_C$ = 21 K, the $V_{out}$ value shows a more significant decrease to 4.40 µV [11]. This suggests that the current sample is quite inhomogeneous and only a small fraction of it becomes SC, although the onset temperature of superconductivity is higher than the other sample. Previous in situ STM and ARPES studies have also shown that the gap size exhibits pronounced variations for different samples, or even for the different areas in the same sample. The inhomogeneous superfluid density can be verified by techniques such as scanning superconducting quantum interference device [18].

Although at the moment we can not give a definitive answer to why this particular sample shows a higher $T_C$, we suspect that it is due to the interface properties between the FeSe film and STO substrate. Previous mutual inductance measurements were performed on FeSe films grown on insulating STO substrate so that the transport properties can be measured simultaneously [11]. However, it is well-known that the surface morphology of the substrate and the charge transfer from the substrate to the FeSe film are very important for the occurrence of high temperature superconductivity in this system. Therefore, it seems quite likely that the optimized interface properties achieved in this Nb doped STO substrate are responsible for the relatively higher $T_C$. This being said, the fact that this sample is more inhomogeneous than the FeSe/STO films with lower $T_C$ suggests that the $T_C$ = 65 K phase here is quite delicate. It is very difficult to achieve or to maintain after capping it with the protection layers for ex situ measurements. We still need to figure out the physics behind the various phases and the way to further optimize the SC properties.

In summary, our mutual inductance measurements provide unambiguous evidence for the onset of Meissner effect at 65 K in FeSe film grown on Nb doped STO substrate by MBE. The onset SC transition temperature is higher than previous record of $T_C$ ~ 55 to 56 K achieved in bulk Sm(O$_{1-x}$F$_x$)FeAs [19], Gd$_{1-x}$Th$_x$FeAsO [20] and SrFeAsF [21]. The FeSe film on Nb doped STO substrate thus has the highest $T_C$ among all iron based superconductors discovered so far. With its simplest crystal structure and highest $T_C$, this system represents the most ideal system for investigating the mechanism of iron based high temperature superconductors. The interface between the iron compounds and oxides also represents a promising playground for finding new superconductors with even higher transition temperature.

## Acknowledgments

This work was supported by the National Natural Science Foundation and Ministry of Science and Technology of China (2015CB921000 and 2012CB921402). Yihua Wang is partially supported by the Urbanek Fellowship of the Department of Applied Physics at Stanford University. K.A. Moler is supported by the Department of Energy, Office of Science, Basic Energy Sciences, Materials Sciences and Engineering Division, under Contract DE-AC02-76SF00515.

**Figure captions:**

Fig. 1 The sample growth and experimental method. **a** Schematic structure of the FeSe film grown on Nb doped STO substrate. **b** Schematic experimental setup of a two-coil mutual inductance measurement system

Fig. 2 Temperature dependence of the Meissner effect of the sample. **a** The out-of-phase voltage $V_{out}$ in the pickup coil measured on the bare substrate (black) and on the FeSe film grown on the substrate (red). **b** The derivative of the two curves shown in (**a**), demonstrating the onset of Meissner effect at around 65 K

Fig.3 Mutual inductance measurements in varied magnetic field. **a** The diamagnetic signal decreases due to the suppression of superconductivity by increasing magnetic field. **b** The $V_{out}$ curves show the broadening of the superconducting transition near the same onset temperature 65 K.

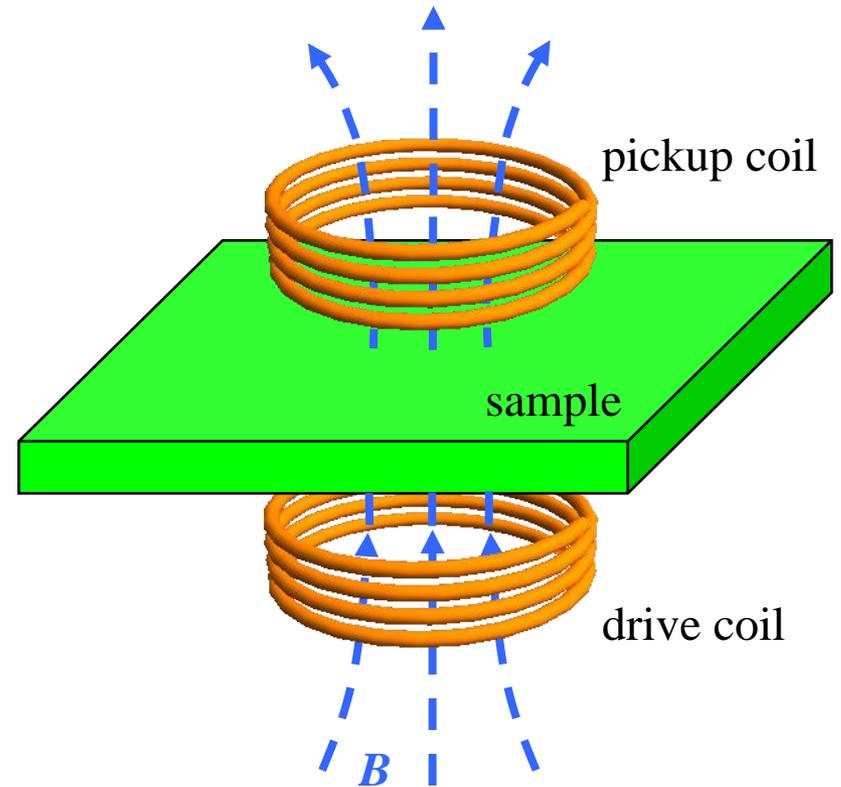

Figure 1

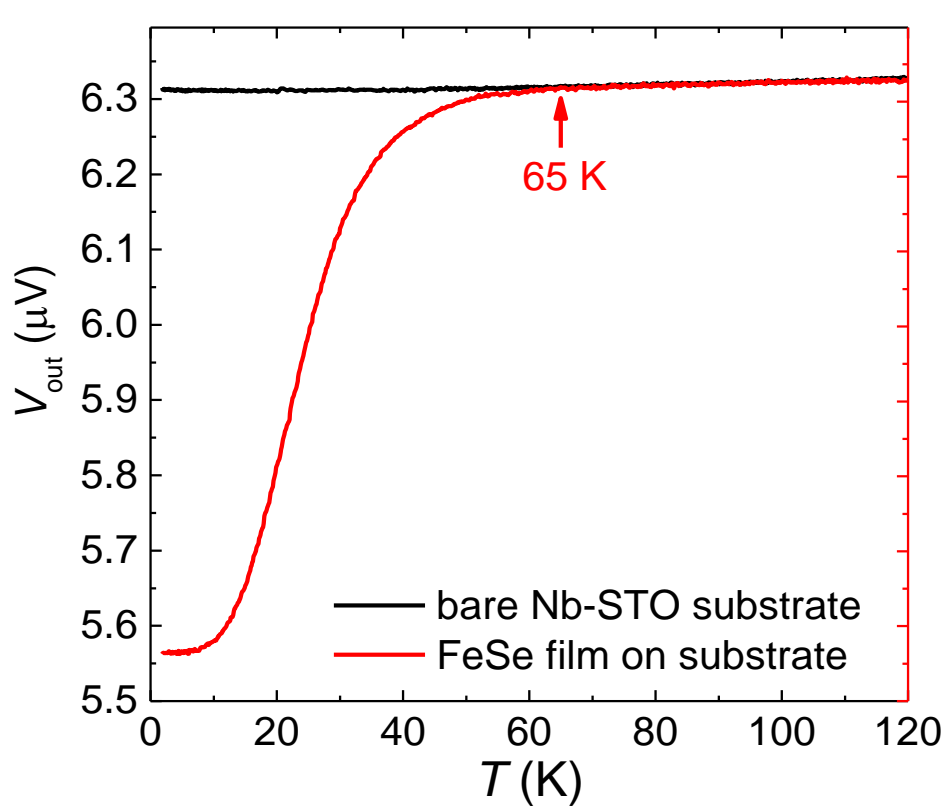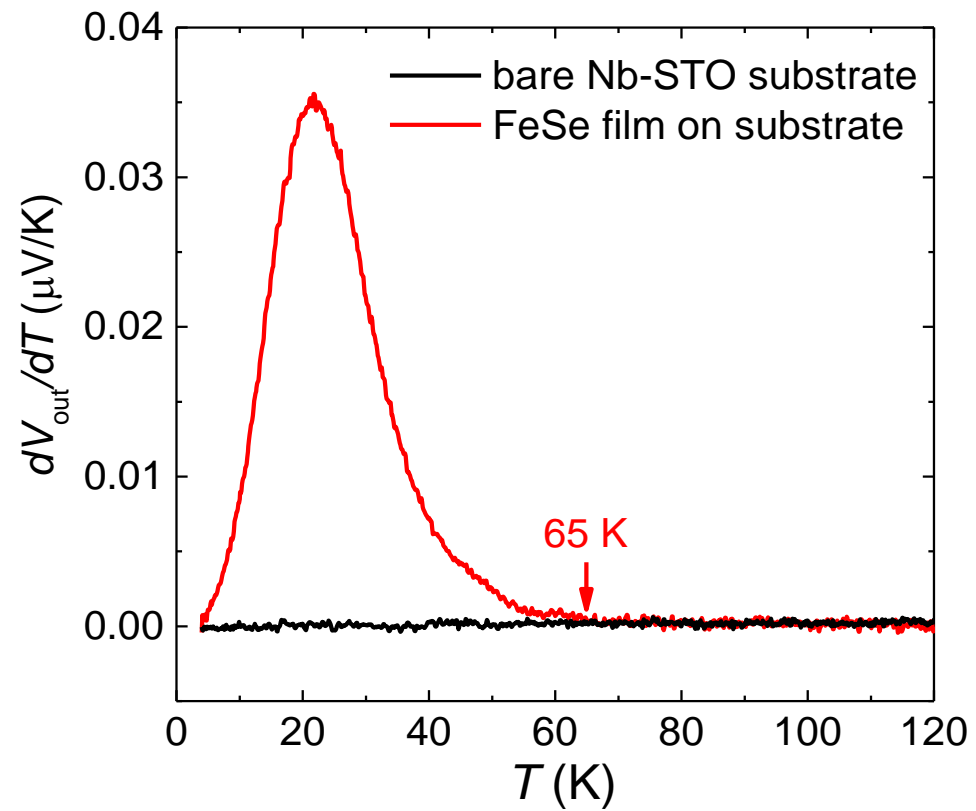

Figure 2

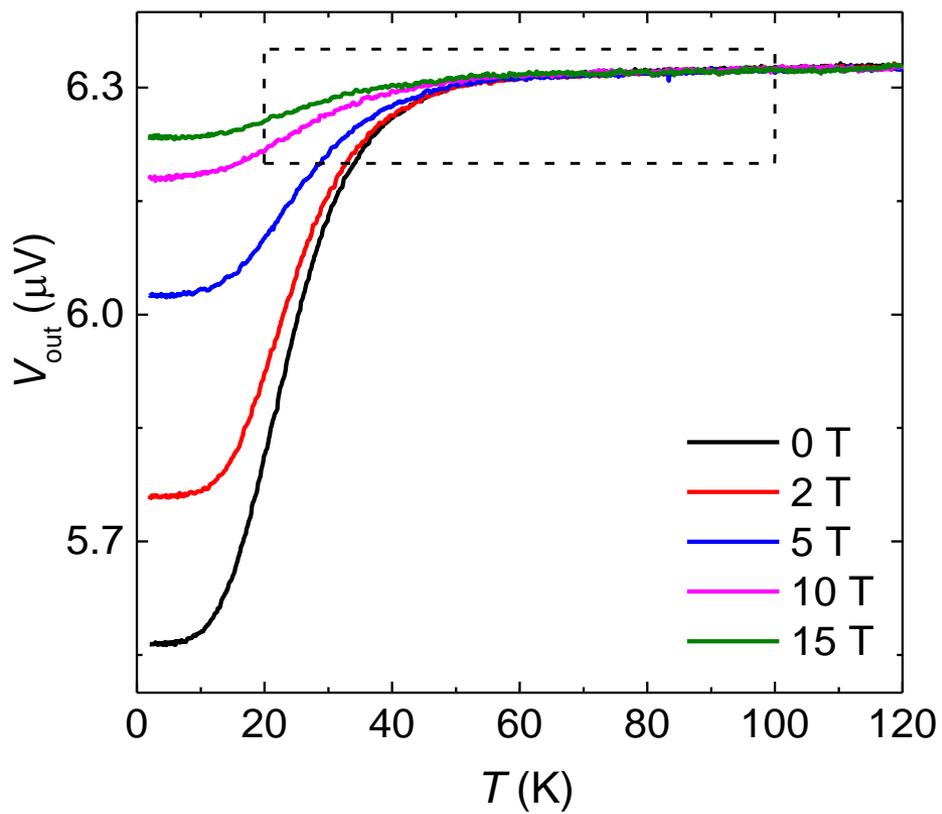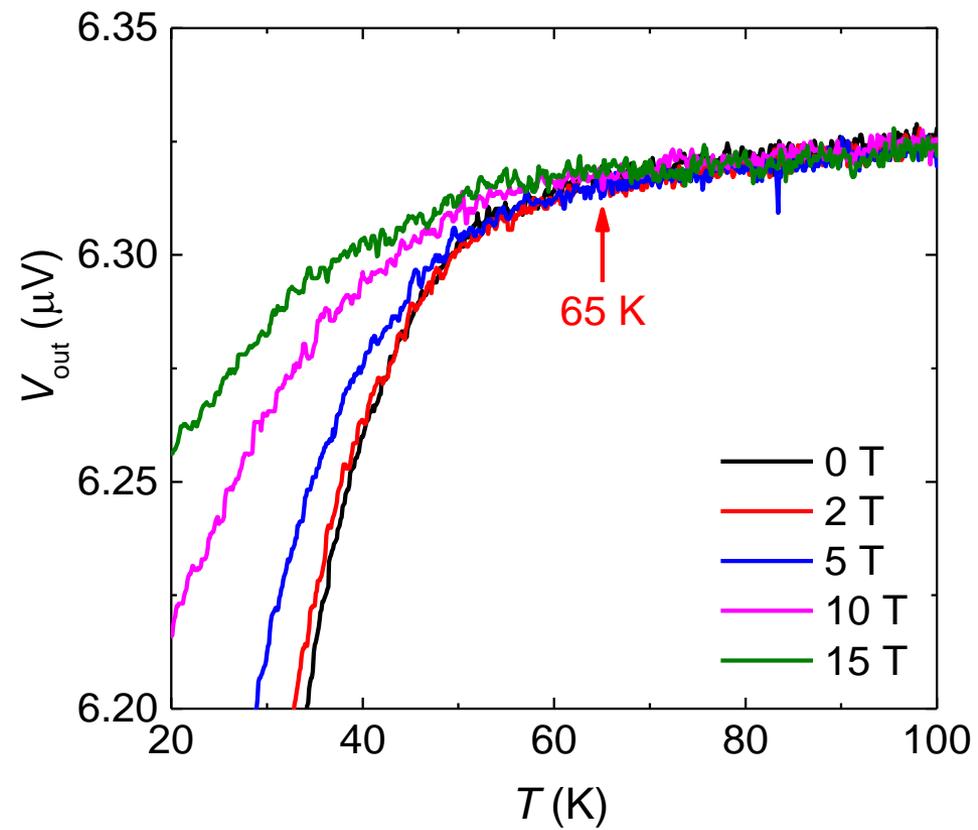

Figure 3